%% file: main.tex
\documentclass[journal,onecolumn,12pt]{IEEEtran}
\usepackage{amssymb}
\usepackage{amsmath}
\usepackage{graphicx}
\usepackage{tcolorbox}

\usepackage[ruled]{algorithm2e}
\usepackage{bm}
\usepackage{romannum}
\usepackage{xcolor}
\usepackage{enumitem}

\usepackage{mathtools}

\makeatletter
\newcommand{\subalign}[1]{%
  \vcenter{%
    \Let@ \restore@math@cr \default@tag
    \baselineskip\fontdimen10 \scriptfont\tw@
    \advance\baselineskip\fontdimen12 \scriptfont\tw@
    \lineskip\thr@@\fontdimen8 \scriptfont\thr@@
    \lineskiplimit\lineskip
    \ialign{\hfil$\m@th\scriptstyle##$&$\m@th\scriptstyle{}##$\hfil\crcr
      #1\crcr
    }%
  }%
}
\makeatother

\newcommand{\norm}[1]{\left\lVert#1\right\rVert}

\DeclareMathOperator*{\argmax}{arg\,max}

\usepackage{siunitx}
\begin{document}
\title{End-to-End Fast Training of Communication Links Without a Channel Model via Online Meta-Learning}

\author{Sangwoo Park,~\IEEEmembership{Student Member,~IEEE,}
        Osvaldo Simeone,~\IEEEmembership{Fellow,~IEEE,}
        and\\~Joonhyuk Kang,~\IEEEmembership{Member,~IEEE}%

\thanks{Code for regenerating the results of this paper can be found at https://github.com/kclip/meta-autoencoder-without-channel-model).}
\thanks{The work of S. Park and J. Kang was supported by the National Research Foundation of Korea (NRF) grant funded by the Korea government (MSIT) (No. 2017R1A2B2012698). The work of O. Simeone was supported by the European Research Council (ERC) under the European Union's Horizon 2020 research and innovation programme (grant agreement No. 725731).}}


\pagenumbering{arabic}

\maketitle
\thispagestyle{plain}
\pagestyle{plain}

\begin{abstract}
When a channel model is not available, the end-to-end training of encoder and decoder on a fading noisy channel generally requires the repeated use of the channel and of a feedback link. An important limitation of the approach is that training should be generally carried out from scratch for each new channel. To cope with this problem, prior works considered joint training over multiple channels with the aim of finding a single pair of encoder and decoder that works well on a class of channels. In this paper, we propose to obviate the limitations of joint training via meta-learning. The proposed approach is based on a meta-training phase in which the online gradient-based meta-learning of the decoder is coupled with the joint training of the encoder via the transmission of pilots and the use of a feedback link. Accounting for channel variations during the meta-training phase, this work demonstrates the advantages of meta-learning in terms of number of pilots as compared to conventional methods when the feedback link is only available for meta-training and not at run time.
\end{abstract}



%
\IEEEpeerreviewmaketitle

\input{intro}

\input{model}
\input{online_set_up.tex}
\input{experiments}

\vspace{-0.2cm}
\input{conclusion}
\small
\input{ack}

\bibliographystyle{IEEEtran}
\bibliography{ref}

\end{document}

%% file: intro.tex
\section{Introduction}
\label{sec:intro}

\begin{figure}[t!]
    \centering
    \includegraphics[width=0.6\columnwidth]{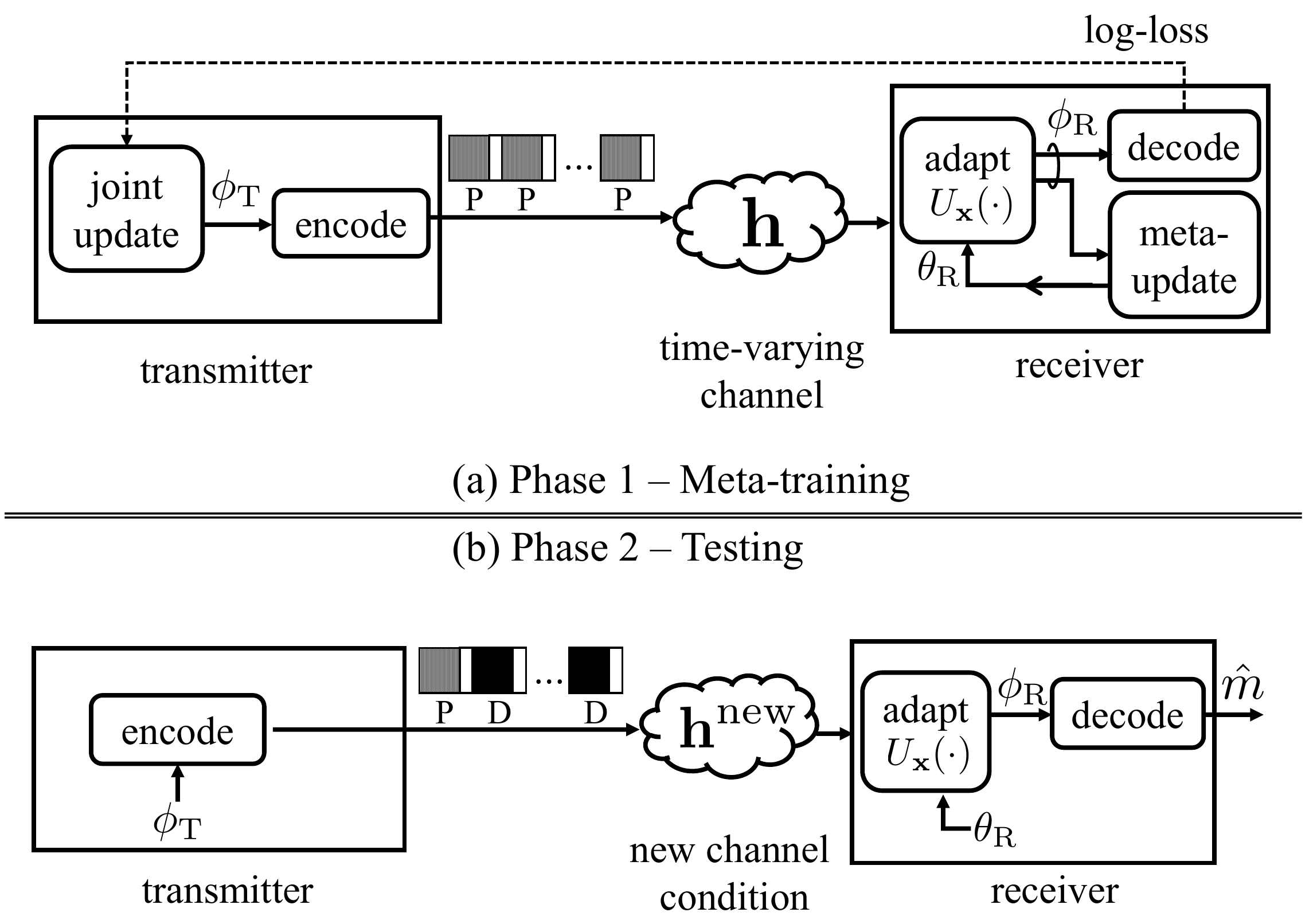}
    \caption{(a) The system first carries out online (meta-)training by transmitting multiple pilot packets (``$\text{P}$") over time-varying channel conditions and by leveraging a feedback link; (b) Then, the system is tested on new channel conditions in the testing phase, which consists of both pilot and data (``$\text{D}$") packets. The feedback link is not available for testing.}
    \label{fig:phases}
\end{figure}

Among the main promises of machine learning for communications is its capability to enable the end-to-end training of a link starting from a blank slate. To fix the ideas, consider an experimental set-up in which a transmitter and a receiver are deployed in a given environment, say in an airport lounge. The transmitter sends a sequence of pilot frames, while the receiver can feedback information to the transmitter on a separate link (see Fig. 1(a)). Pilot and feedback signals are used to train encoder and decoder in an end-to-end fashion in the absence of a channel model. After this training phase, the link is operated at run time by following a conventional frame-based transmission protocol, which encompasses the transmission of both pilots and data payload and does not allow for feedback (see Fig.~\ref{fig:phases}(b)). The goal is for the link to meet quality of service requirements, typically in terms of error rate, during deployment, also referred to as testing.

A key challenge in meeting this design goal is the need for the link to generalize its operation from the channel conditions experienced during training to the a priori unknown conditions to be encountered during testing. 
A standard approach would be to train by continuously adapting encoder and decoder as more pilots are collected during the training phase. To this end, one can implement the state-of-the-art approach proposed in \cite{aoudia2019model} in which the decoder is trained via supervised learning, while the encoder is trained using policy gradient-based reinforcement learning on the basis of feedback from the receiver. By continuously running this training algorithm, one would effectively train a single encoder and decoder pair that optimizes an average performance criterion across the experienced channel conditions. 

The conventional approach outlined above, which we refer to as \emph{joint training}, has the apparent drawback that there may not be a single pair of encoder and decoder that guarantees a desirable average performance. In fact, joint training effectively approximates a non-coherent transmission and reception solution. An improvement may be obtained by adapting the jointly trained decoder by using the pilots received during testing in each frame (Fig.~\ref{fig:phases}(b)). However, joint training does not cater explicitly for adaptation to new channel conditions, and the performance may not exceed that of conventional training from scratch \cite{park2019learning_conf, park2019learning}.

Based on the presence of pilots in each frame, a more relevant learning goal during the training phase is, therefore, that of inferring an effective decoder's adaptation rule based on the received pilots during deployment, while jointly training the encoder. Note that the encoder cannot adapt during deployment due to the assumed lack of a feedback link at run time. The jointly trained encoder and the decoder's adaptation rule should guarantee an averaged optimized performance by accounting for the fact that the decoder is adapted using pilots in the current frame.

Using the most common terminology (see review in \cite{simeone2020learning}), the proposed approach carries out the joint training of the encoder and the meta-training of the decoder by using observations from multiple channel realizations during the training phase (Fig.~\ref{fig:phases}(a)). Meta-learning, or learning to learn, refers here to the training of the adaptation rule, and, more generally, to the inference of an inductive bias \cite{simeone2020learning}. We specifically build on the training algorithm in \cite{aoudia2019model} mentioned above, which is suitably integrated with the Model-Agnostic Meta-Learning (MAML) algorithm \cite{finn2017model}. MAML was previously applied to the end-to-end training of links in the presence of a channel model -- unlike the set-up considered here -- in \cite{park2019meta}. Other applications of MAML to communication systems include the meta-training of demodulator or decoder for a fixed encoder \cite{park2019learning, jiang2019mind}, channel estimation \cite{mao2019roemnet}, and uplink/downlink channel state information conversion \cite{yang2019deep}. The use of alternative meta-training algorithms, such as fast Context Adaptation VIA meta-learning (CAVIA) \cite{zintgraf2018fast} and REPTILE \cite{nichol2018first} is also explored in \cite{park2019learning}.

%% file: model.tex
\section{System Model}
\label{sec:model}

\begin{figure}[t!]
    \centering
    \includegraphics[width=0.6\columnwidth]{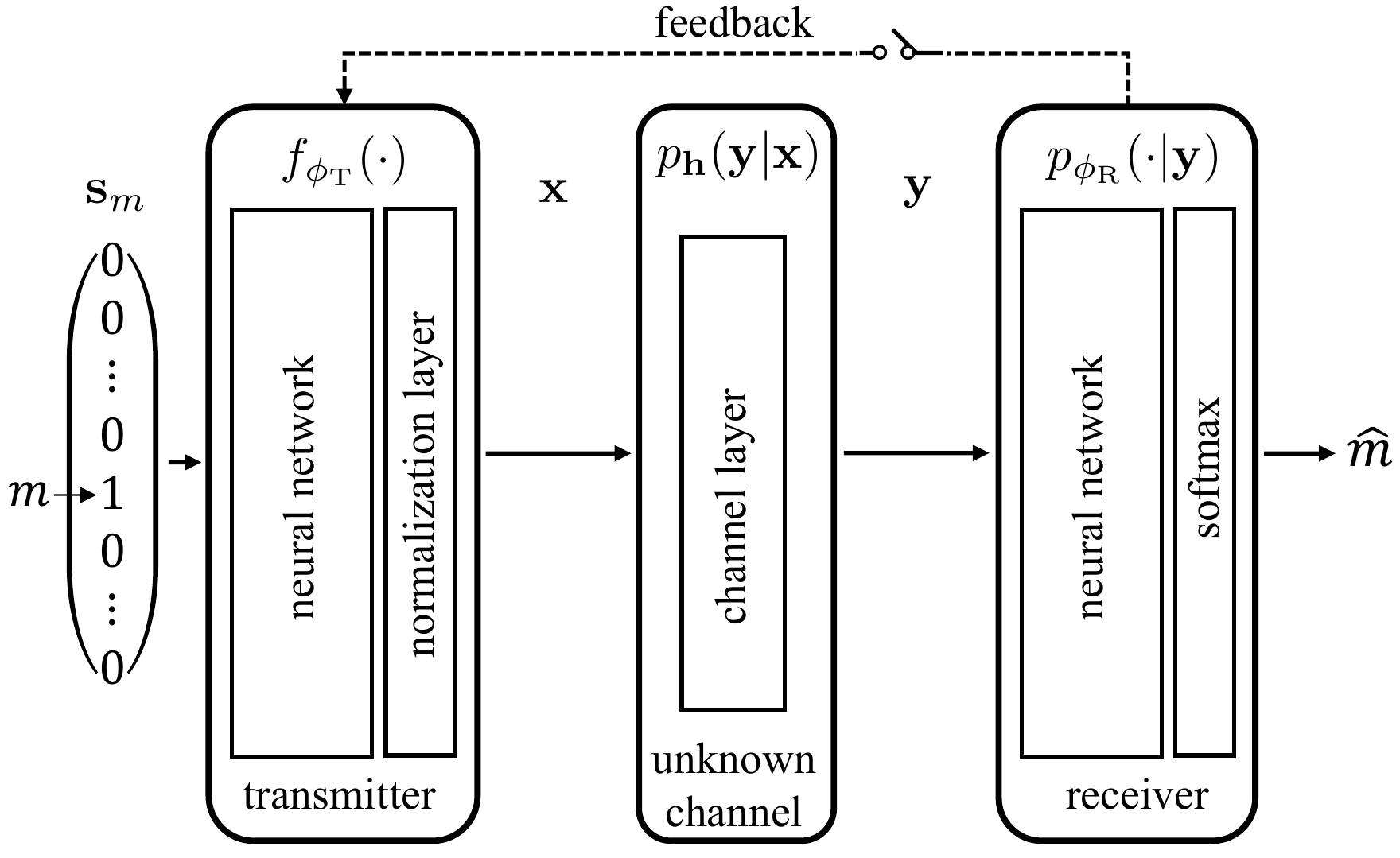}
    \caption{Training a link without a channel model: A message $m$ is mapped into a codeword $\mathbf{x}$ via a trainable encoder $f_{\phi_\text{T}}(\cdot)$, while the received signal $\mathbf{y}$, determined by the unknown channel $p_{\textbf{h}}(\mathbf{y}|\mathbf{x})$, is mapped into estimated message $\hat{m}$ through a trainable decoder $p_{\phi_\text{R}}(\cdot|\mathbf{y})$. A feedback link is available between transmitter and receiver during the (meta-)training phase (Fig.~\ref{fig:phases}(a)) but not during testing (Fig.~\ref{fig:phases}(b)).}
    \label{fig:model}
\end{figure}

We consider the set-up of \cite{aoudia2019model}, in which the goal is to train in an end-to-end fashion encoder and decoder of a communication system in the absence of a channel model. As illustrated in Fig.~\ref{fig:model}, encoder and decoder are modeled as neural networks with trainable weights $\phi_\text{T}$ and $\phi_\text{R}$, respectively.

As shown in Fig.~\ref{fig:phases}, we consider a two-phase operation. In the (meta-)training phase (Fig.~\ref{fig:phases}(a)), the transmitter sends multiple pilot (``P") blocks, experiencing generally correlated channel realizations. In order to facilitate generalization to new channel conditions, in practice, the channel can be made to vary by modifying the position of transmitter and receiver or by changing the propagation environment, e.g., by moving objects around transmitter and receiver. In the second, deployment or testing phase (Fig.~\ref{fig:phases}(b)), the performance of the link is tested on new channel conditions. In this phase, the transmitter sends pilot blocks followed by data (``$\text{D}$") blocks, and the feedback link is disabled.


In each transmission block $t=1,2,\ldots,$ be it a pilot or data block, the encoder takes as input a one-hot vector $\mathbf{s}_m(t)$ of dimension $2^k$, which represents a message $m(t)\in\{1,\ldots,2^k\}$ of $k$ bits. Vector $\mathbf{s}_m(t)$ has a single entry equal to ``1'' in position $m(t)$, with all other entries equal to zero. The encoder maps each input $\mathbf{s}_m(t)$ into a transmitted vector $\mathbf{x}(t)\in\mathbb{C}^n$ of $n$ complex symbols or, equivalently, $2n$ real symbols. As seen in Fig.~\ref{fig:model}, the encoding from $\mathbf{s}_m(t)$ to $\mathbf{x}(t)$ is done through a trainable mapping $\mathbf{x}(t)=f_{\phi_\text{T}}(\mathbf{s}_m(t))$, which is defined by a neural network with weight vector $\phi_\text{T}$ and by a normalization layer that ensures the total power constraint $\norm{\mathbf{x}(t)}^2/n=E_s$.

The codeword $\mathbf{x}(t)$ is transmitted through a channel, whose model  $p_{\mathbf{h}(t)}(\mathbf{y}|\mathbf{x})$ is unknown to both transmitter and receiver, to produce the received signal $\mathbf{y}(t)$. We will specifically assume the general model 
\begin{align} \label{eq:channel_model}
\mathbf{y}(t) = \mathbf{h}(t)*\mathbf{x}(t) + \mathbf{w}(t),
\end{align}
where $\mathbf{w}(t)\sim\mathcal{CN}(0,N_0\mathbf{I})$ represents complex Gaussian i.i.d. noise and ``$*$'' indicates a linear operation on input $\mathbf{x}(t)$ parameterized by a channel vector $\mathbf{h}(t)$. This accounts for a convolution for time-domain transmission and to entry-wise multiplication for frequency-domain transmission.

The receiver passes the received signal through a neural network parameterized by a weight vector $\phi_\text{R}$ that outputs a $2^k\times 1$ vector of probabilities $p_{\phi_\text{R}}(m|\mathbf{y}(t))$ for $m\in\{ 1,\ldots,2^k \}$. Each output $p_{\phi_\text{R}}(m|\mathbf{y}(t))$ provides an estimate of the corresponding posterior probability that the transmitted message is $m(t) = m$. A final hard estimate can be obtained via the approximate maximum a posteriori (MAP) rule
$\hat{m}(t) =\argmax_{m=1,\ldots,2^k} p_{\phi_\text{R}}(m|\mathbf{y}(t)).$ 

The $L\times1$ channel vector $\mathbf{h}(t)$ remains constant for frames of $T$ transmission blocks, and it changes according to a correlated process across different frames. We specifically assume the autoregressive Rayleigh fading process
\begin{align} \label{eq:ar_model}
\mathbf{h}(t) = \begin{cases}\rho \mathbf{h}(t-T) + \sqrt{1-\rho^2} \tilde{\mathbf{h}}(t), &\text{ if } t = nT, \\
\mathbf{h}(t-1), &\text{ if } t \neq nT,
\end{cases}
\end{align}
for $n=1,2,\ldots,$ where $0\leq\rho\leq1$ is the correlation coefficient and $\tilde{\mathbf{h}}(t)\sim \mathcal{CN}(0, L^{-1}\mathbf{I}_L)$ is the innovation term, which is independent of all other random variables. The training phase takes place across multiple frames of $T$ blocks. For testing, pilots and data payload are transmitted within the same frame of $T$ blocks.

%% file: online_set_up.tex
\section{Online Hybrid Joint and Meta-Training}
\label{sec:online_meta_learning}

In this section, we describe the operation of the proposed system during the (meta-)training phase. As discussed in Sec.~\ref{sec:intro}, a conventional joint training strategy would train a single encoder and decoder pair to perform well on average over the distribution of the channels observed during this phase. The drawback of this approach is that it does not account for the possibility to update the decoder based on pilot symbols present in each frame during testing. The proposed scheme tackles this problem by jointly training the encoder and meta-training the decoder. The receiver meta-trains the decoder to quickly adapt to new channel condition based on the pilots preceding the payload in each frame; while the transmitter simultaneously trains a single encoder. Training of encoder and decoder is carried out at the same time, enabling the jointly trained encoder to choose a codebook that is tailored to a ``coherent'' decoder adapted to the current channel based on the pilots.

In this section, we use $\tau = 1,2,\ldots,$ to index the frames in the training phase, and write $[\tau]=((\tau-1)T+1,\ldots,\tau T)$ to denote the set of block indices in the $\tau$th frame. Accordingly, we also write $\mathbf{m}[\tau] = \{ m(t): t \in [\tau] \}$ to denote the $T$ messages sent in the $\tau$th frame, and we similarly introduce the channel $\mathbf{h}[\tau] = \{ \mathbf{h}(t): t \in [\tau] \}$, the transmitted symbols $\mathbf{x}[\tau] = \{ \mathbf{x}(t): t \in [\tau] \}$, and the received symbols $\mathbf{y}[\tau] = \{ \mathbf{y}(t): t \in [\tau] \}$ in the $\tau$th frame. 

Based on a subset $T_U\leq T$ of blocks in each frame $[\tau]$, the receiver updates the decoder parameter vector as $\phi_\text{R}=U_{\mathbf{x}[\tau_U]}(\theta_\text{R})$, where $[\tau_U]$ denotes the indices of the $T_U$ pilots used to update the decoder. Starting from an initialization parameter $\theta_\text{R}$, function $U_{\mathbf{x}[\tau_U]}(\theta_\text{R})$ produces an adapted decoder parameter $\phi_\text{R}$ based on the pilots $\mathbf{x}[\tau_U]$. 

As in \cite{aoudia2019model}, we adopt the stochastic encoder $\mathbf{x}(t) \sim \pi_{\phi_\text{T}}(\cdot|m(t)) = \mathcal{N}(\sqrt{ 1-\sigma^2 } f_{\phi_\text{T}}(\mathbf{s}_{m}(t)), \sigma^2 \mathbf{I})$ for some $\sigma^2$ \cite{aoudia2019model}. Adding Gaussian noise to the encoded signal $f_{\phi_\text{T}}(\mathbf{s}_{m}(t))$ enables the exploration of the space of the transmitted signals. Note that the original deterministic encoder is obtained as a special case when $\sigma^2=0$.

The objective of the training phase is to (approximately) minimize the average cross-entropy loss per frame
\begin{align} \label{eq:meta-goal-final}
\min_{\phi_\text{T}, \theta_\text{R}} L(\phi_\text{R},\theta_\text{R}) \dot{=}  \mathbb{E}_{\subalign{\mathbf{h}[\tau]&\sim p_{\mathbf{h}}(\cdot), \\\mathbf{m}[\tau] &\sim p_{\mathbf{m}}(\cdot), \\\mathbf{x}[\tau] &\sim \pi_{\phi_\text{T}}(\cdot|\mathbf{m}[\tau]) \\ \mathbf{y}[\tau] &\sim p_{\mathbf{h}[\tau]} (\cdot|\mathbf{x}[\tau])}}\left[-\sum_{t\in[\tau]}\log p_{\phi_\text{R}=U_{\mathbf{x}[\tau_U]}(\theta_\text{R})} (m(t)|\mathbf{y}(t))\right],
\end{align}
which is averaged over the channel $\mathbf{h}[\tau]\sim p_{\mathbf{h}}(\cdot)$ and over distributions of i.i.d. messages $\{m(t)\sim p_\mathbf{m}(\cdot)\}_{t\in[\tau]}$, codewords $\{ x(t) \sim \pi_{\phi_\text{T}}(\cdot|m(t)) \}_{t\in[\tau]}$, and received signals $\{ y(t) \sim p_{\mathbf{h}(t)}(\cdot|x(t)) \}_{t\in[\tau_U]}$.

\subsubsection{Joint Training of the Encoder}
The gradient with respect to $\phi_\text{T}$ of the objective function of problem \eqref{eq:meta-goal-final} can be written as (see, e.g., \cite{simeone2018brief})
\begin{align}\label{eq:grad_tx}
\nabla_{\phi_\text{T}} L(\phi_{\text{T}},\theta_{\text{R}}) = \mathbb{E}_{\subalign{\mathbf{h}[\tau]&\sim p_{\mathbf{h}}(\cdot),\\
\mathbf{m}[\tau] &\sim p_\mathbf{m}(\cdot), \\\mathbf{x}[\tau] &\sim \pi_{\phi_\text{T}}(\cdot|\mathbf{m}[\tau]) \\ \mathbf{y}[\tau] &\sim p_{\mathbf{h}[\tau]} (\cdot|\mathbf{x}[\tau])}} &\left[-\sum_{t \in [\tau]}\log p_{U_{\mathbf{x}[\tau_U]}(\theta_\text{R})} (m(t)|\mathbf{y}(t)) 
\nabla_{\phi_\text{T}} \log \pi_{\phi_\text{T}}(\mathbf{x}(t)|m(t))\right].
\end{align}
We emphasize that the same distribution $\pi_{\phi_\text{T}}(\cdot|m(t))$ is used to generate pilots used by the receiver to adapt the decoder and to define the codewords for data transmission. This assumption may be alleviated in future work.
In order to train the encoder, one can estimate the gradient \eqref{eq:grad_tx} via an empirical average based on the received signals as
\begin{align} \label{eq:grad_tx_emp}
\hat{\nabla}&_{\phi_\text{T}}L(\phi_{\text{T}},\theta_{\text{R}})= \frac{1}{F} \sum_{\tau = 1}^{F} \frac{1}{T}\sum_{t \in [\tau]}\Bigg[-\log p_{U_{\mathbf{x}[\tau_U]}(\theta_\text{R})} (m(t)|\textbf{y}(t)) 
\nabla_{\phi_\text{T}}  \log \pi_{\phi_\text{T}}(\textbf{x}(t)|m(t))\Bigg],
\end{align}
where $F$ is the number of frames transmitted in the training phase. The proposed scheme applies the Stochastic Gradient Descent (SGD) update
\begin{align} \label{eq:joint_tx_update_rule}
\phi_\text{T}^{(\tau+1)} &\leftarrow \phi_\text{T}^{(\tau)} - \frac{\kappa}{T}\sum_{t \in [\tau]}\Bigg[-\log p_{U_{\mathbf{x}[\tau_U]}(\theta_\text{R})} (m(t)|\textbf{y}(t)) \nabla_{\phi_\text{T}}\log \pi_{\phi_\text{T}}(\textbf{x}(t)|m(t))\Bigg],
\end{align}
for some learning rate $\kappa>0$ across the training frames $\tau=1,2,\ldots,F$. In order to enable the evaluation of \eqref{eq:joint_tx_update_rule}, the log-loss $-\log p_{U_{\mathbf{x}[\tau_U]}(\theta_\text{R})} (m(t)|\mathbf{y}(t))$, for all $t\in[\tau]$, is communicated by the receiver via the noiseless feedback link at the end of each frame $\tau$ in the (meta-)training phase as shown in Fig.~\ref{fig:phases}(a).

\subsubsection{Meta-Learning of the Decoder}
\label{subsec:meta_learning}
The receiver updates the decoder via SGD, yielding the update function 
\begin{align} \label{eq:meta_local_update}
\phi_\text{R} = U_{\mathbf{x}[\tau_U]}(\theta_\text{R}) = \theta_\text{R} - \frac{\eta}{T}\sum_{t' \in [\tau_U]}[-\nabla_{\theta_\text{R}}\log p_{\theta_\text{R}} (m(t')|\mathbf{y}(t'))],
\end{align}
for some learning rate $\eta>0$. The update function \eqref{eq:meta_local_update} can also be generalized to allow for a larger number of SGD updates, rather than a single one as in \eqref{eq:meta_local_update}.

With this choice for the update function, the gradient of objective function of problem \eqref{eq:meta-goal-final} with respect to $\theta_\text{R}$ can be written as (see, e.g., \cite{simeone2018brief})
\begin{align}\label{eq:grad_rx}
\nonumber
\nabla_{\theta_\text{R}} L(\phi_{\text{T}},\theta_{\text{R}}) &= \mathbb{E}_{\subalign{\mathbf{h}[\tau]&\sim p_{\mathbf{h}}(\cdot)\\
\mathbf{m}[\tau] &\sim p_\mathbf{m}(\cdot), \\\mathbf{x}[\tau] &\sim \pi_{\phi_\text{T}}(\cdot|\mathbf{m}[\tau]) \\ \mathbf{y}[\tau] &\sim p_{\mathbf{h}[\tau]} (\cdot|\mathbf{x}[\tau])}} \left[-\sum_{t\in[\tau]}\nabla_{\theta_\text{R}}\log p_{U_{\mathbf{x}[\tau_U]}(\theta_\text{R})} (m(t)|\mathbf{y}(t))\right]
\nonumber\\
&=\mathbb{E}\Bigg[-\sum_{t \in [\tau]}\mathbf{J}_{\theta_\text{R}}U_{\mathbf{x}[\tau_U]}(\theta_\text{R})  \nabla_{\phi_\text{R}}\log p_{\phi_\text{R}} (m(t)|\mathbf{y}(t))\Bigg] \nonumber\\
&= \mathbb{E}\Bigg[-\sum_{t \in [\tau]}\bigg[\Bigg(I+\frac{\eta}{T}\bigg( \sum_{t' \in [\tau_U]}\nabla^2_{\theta_\text{R}}\log p_{\theta_\text{R}} (m(t')|\mathbf{y}(t'))\bigg)\Bigg) \nabla_{\phi_\text{R}}\log p_{\phi_\text{R}} (m(t)|\mathbf{y}(t))\bigg]\Bigg],
\end{align}
with $\phi_\text{R}$ in \eqref{eq:meta_local_update} and $\mathbf{J}_{\theta_\text{R}}$ representing the Jacobian operator. Finally, the gradient \eqref{eq:grad_rx} can be estimated using the received $F$ frames as in \eqref{eq:grad_tx_emp} producing estimate $\hat{\nabla}_{\theta_\text{R}}L(\phi_\text{T},\theta_\text{R})$, yielding the SGD update rule 
\begin{align}\label{eq:meta-online-update}
\theta_\text{R}^{(\tau+1)} \leftarrow \theta_\text{R}^{(\tau)} - \kappa 
\hat{\nabla}_{\theta_\text{R}}L(\phi_\text{T},\theta_\text{R}).
\end{align}

%% file: experiments.tex
\section{Experiments}
\label{sec:exp}
\label{subsec:exp-conv}
In this section, we provide numerical results to benchmark the performance of the proposed hybrid joint and meta-training neural encoder and decoder against: \emph{(i)} Binary Phase-Shift Keying (BPSK) transmitter with maximum likelihood decoder and Minimum Mean Square Error (MMSE) channel estimation obtained using the pilots in each frame of the testing phase; \emph{(ii)} BPSK transmitter with neural decoder, which is either trained from scratch using the pilots in each frame of the testing phase, or else jointly or meta-trained during the training phase; and \emph{(iii)}
jointly trained neural encoder and decoder.

All experiments assume $k=8$ bits transmission with $n=8$ complex channel uses under an $L=3$-tap Rayleigh fading channel.  The signal-to-noise ratio (SNR) is given as $E_s/N_0=10\text{ dB}$, which corresponds to an outage probability of $0.01$. 
All schemes are tested using the same number $P$ of pilot blocks in frame during testing phase and the performance is measured in terms of block error rate (BLER) on payload data. We assume that each block has sufficiently long guard interval, so that intersymbol interference is limited within each block. During the training phase, we used learning rates $\kappa=0.01$ and $\eta=0.1$ with standard deviation for stochastic encoder $\sigma=0.15$; while during the testing phase the learning rate $\eta=0.1$ is used for meta-training schemes and $\eta=0.001$ for other schemes with a deterministic encoder $(\sigma=0)$. 
The neural encoder and neural decoder, parameterized by $\phi_\text{T}$ and $\phi_\text{R}$, respectively, have the same structure introduced in \cite{o2017introduction} with decoder being equipped with a Radio Transformer Networks (RTN).

During (meta-)training, each frame consists of $T=256$ transmission blocks that include all $2^k=256$ messages. For adaptation, we use $T_U=8$ blocks and $1$ adaptation step in \eqref{eq:meta_local_update}. For testing, the $P$ pilot blocks in each frame are randomly selected among $256$ messages without replacement and the update in \eqref{eq:meta_local_update} is carried out with $T_U=P$ blocks. The BLER is calculated with $10^4$ payload data blocks.

\subsection{Convergence}
\begin{figure}[t]
    \flushleft
    \hspace*{-0.2in}
    \centering
    \includegraphics[width=0.6
    \columnwidth]{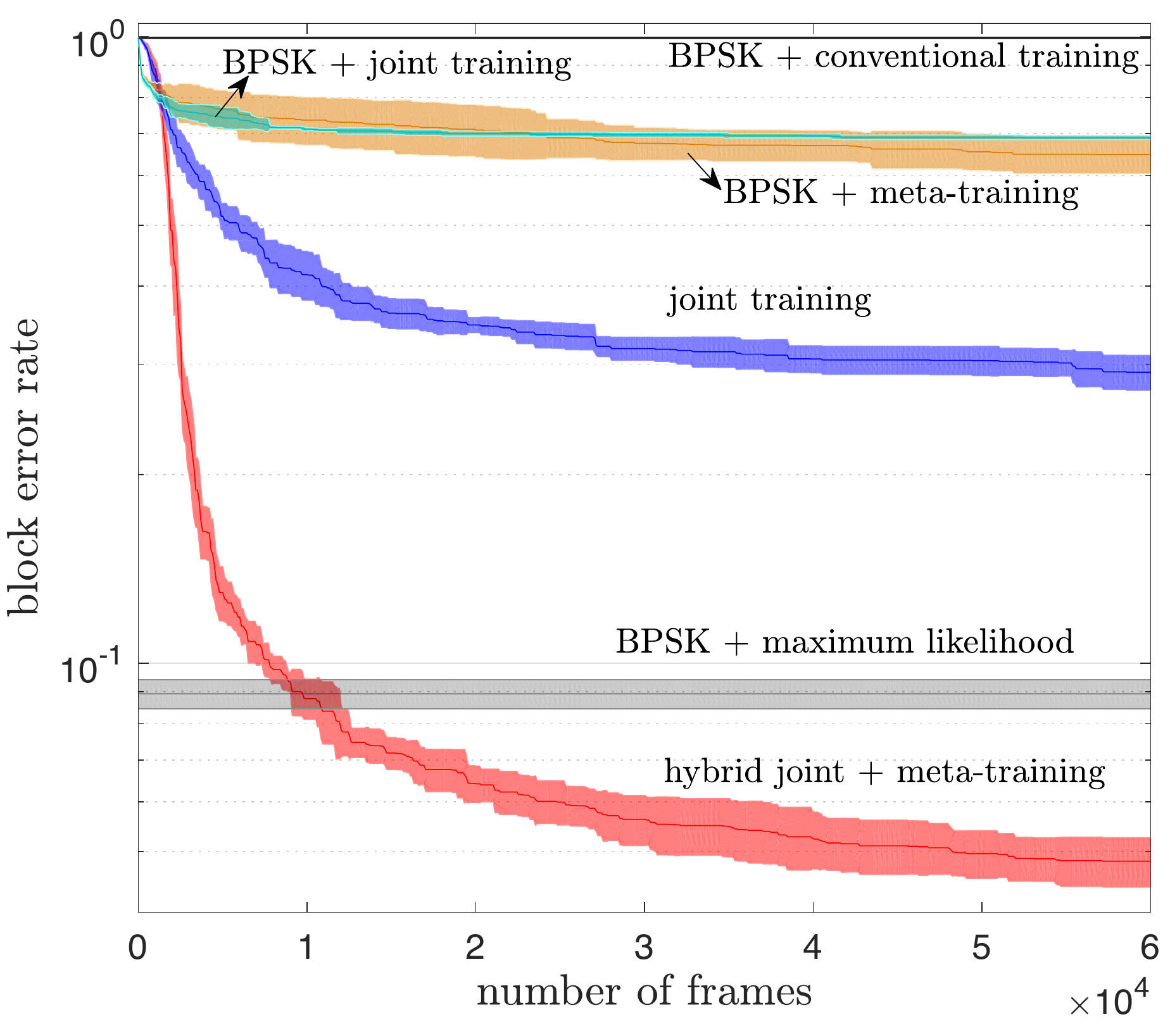}
    \caption{Block error rate (BLER) for new channels during testing as a function of the number of frames used during training phase ($k=8$ bits, $n=8$ complex channel uses, $P=8$ pilot blocks during testing phase; mean and error bars for one standard deviation are shown over $5$ independent experiments).}
    \label{fig:exp_1}
\end{figure}

First, we plot the BLER for the best available solution so far for different schemes as a function of the number of frames during (meta-)training. As it can be seen in Fig.~\ref{fig:exp_1}, owing to the low number $P$ of pilots in each testing frame, only the proposed hybrid scheme is able to outperform BPSK with maximum likelihood decoder under MMSE channel estimation. This gain shows not only that the meta-trained decoder can adapt well to the new channel conditions, but also that the jointly trained encoder is able to facilitate decoder adaptation while also yielding a good codebook for data transmission.

\subsection{Number of Pilots}
\begin{figure}[t]
    \flushleft
    \hspace*{-0.2in}
    \centering
    \includegraphics[width=0.6
    \columnwidth]{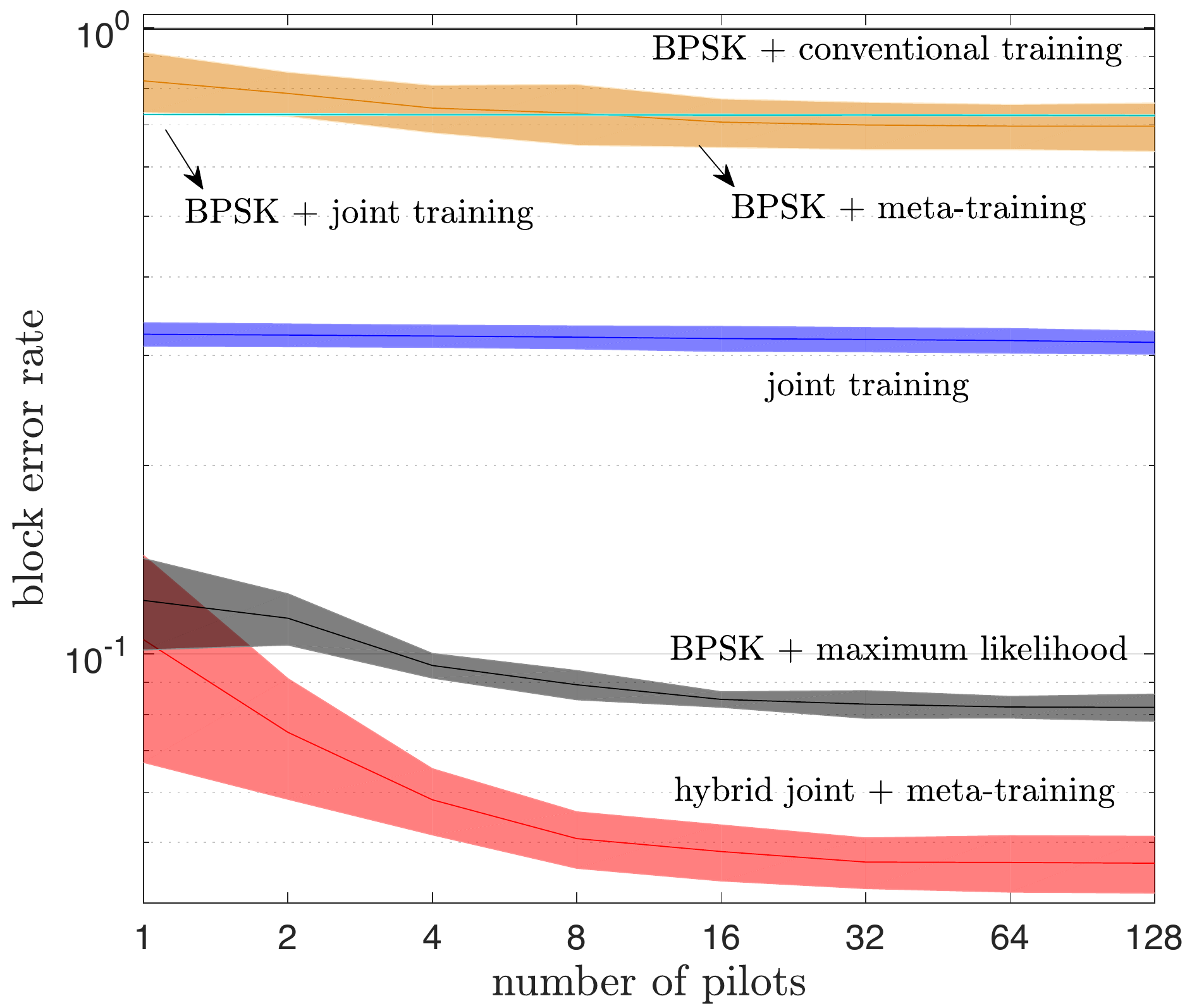}
    \caption{Block error rate (BLER) for new channels during testing as a function of the number of pilot blocks $P$ in testing phase ($k=8$ bits, $n=8$ complex channel uses; mean and error bars for one standard deviation are shown over $5$ independent experiments).}
    \vspace{-0.2cm}
    \label{fig:exp_2}
\end{figure}

Using the best performing solution for each scheme after $60,000$ iterations in the (meta-)training phase, Fig.~\ref{fig:exp_2} elaborates on the capability of different strategies to adapt to current channel condition by plotting the BLER versus the number $P$ of pilots in each testing frame. The figure shows that BPSK transmitter with maximum likelihood decoder generally shows best performance if there is only one pilot block $(P=1)$, while, with more pilots, the proposed hybrid scheme outperforms all the other schemes. This follows the intuition that model-based schemes, here BPSK with maximum likelihood decoder and MMSE channel estimation, can work with fewer pilots; while data-based schemes based on meta-training can significantly improve the performance when the number of pilots is sufficiently large (here $P>1$).

\subsection{Channel Correlation}
\label{subsec:exp-corr}
\begin{figure}[t]
    \flushleft
    \hspace*{-0.2in}
    \centering
    \includegraphics[width=0.6
    \columnwidth]{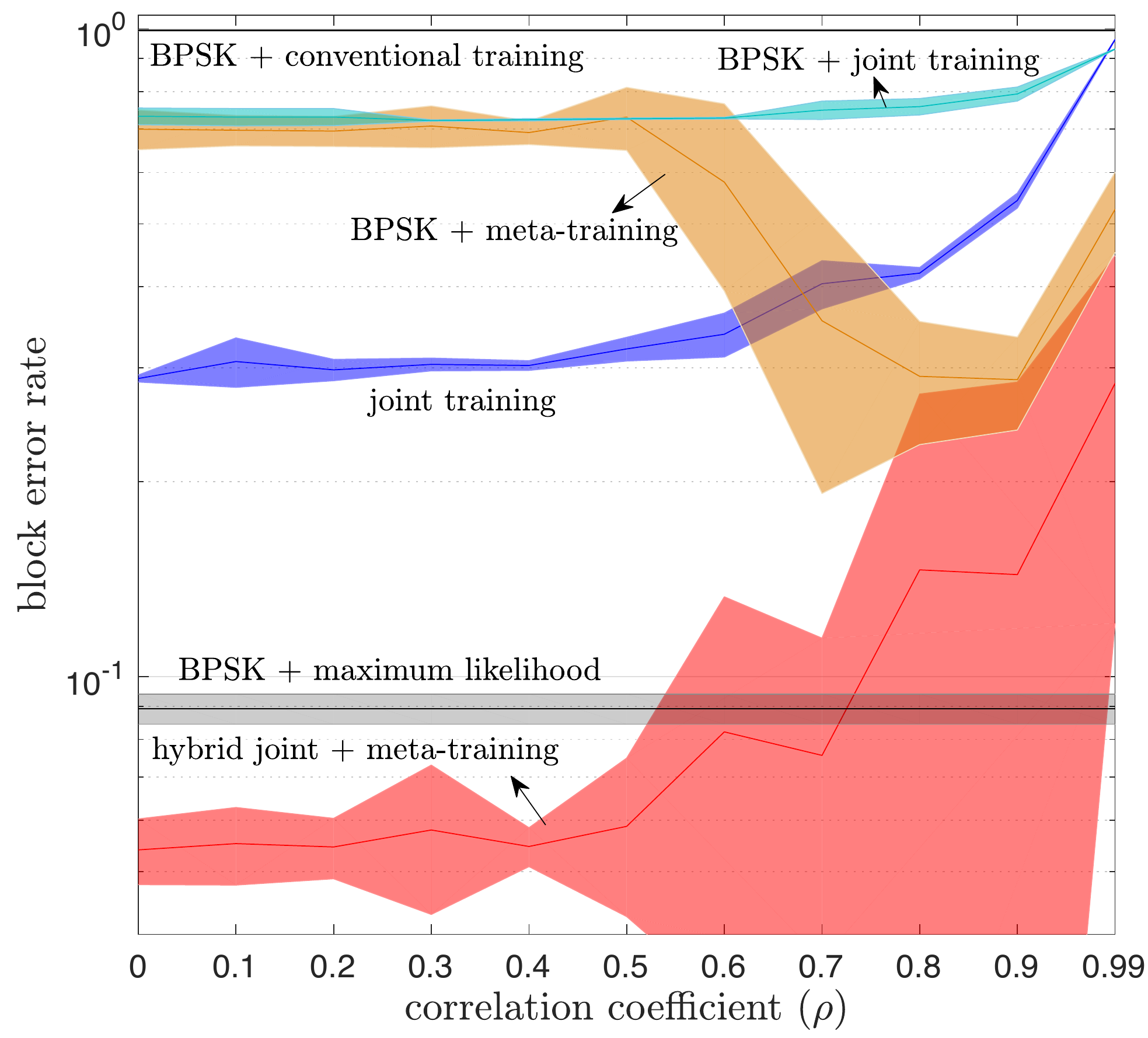}
    \caption{Block error rate (BLER) for new channels during testing as a function of the correlation coefficient $\rho$ in \eqref{eq:channel_model} ($k=8$ bits, $n=8$ complex channel uses, $P=8$ pilot blocks during testing phase; mean and error bars for one standard deviation are shown over $15$ independent experiments for proposed hybrid scheme while $5$ experiments for other schemes).}
    \vspace{-0.6cm}
    \label{fig:exp_3}
\end{figure}

Finally, we analyze the impact of the channel correlation $\rho$ in \eqref{eq:channel_model} during the training phase. While conventional schemes are seen to be generally impaired by an increased correlation, this may not be the case for methods that rely on meta-learning. On the one hand, meta-learning needs to be exposed to different channels, hence benefiting from a smaller $\rho$. In fact, a large $\rho$ may cause meta-overfitting due to the similarity of the channels observed during meta-training, making adaptation to new channel condition potentially less effective. Meta-overfitting is discussed for general applications in \cite{yin2019meta, rothfuss2020pacoh}. This phenomenon is reflected in the large variability of the BLER in Fig.~\ref{fig:exp_3}, which demonstrates that the meta-trained decoder may not generalize well to all channel realizations when $\rho$ is large. On the other hand, an excessively small value of $\rho$ increases the variance of the meta-training updates \eqref{eq:grad_tx}--\eqref{eq:meta-online-update} by reducing the number of samples available for each channel realization.

%% file: conclusion.tex
\section{Concluding Remarks}
\label{sec:conclusion}
This paper considered a set-up in which a link, exposed to varying channel conditions, learns an encoder and a decoder's adaptation rule based on pilots received in each frame. The proposed approach, which leverages meta-learning, was seen to significantly reduce the number of pilots needed in each frame. Future work may consider adapting the transmitter via feedback during run time, and carrying out experimental tests using software-defined radio.

%% file: ack.tex
\section{Acknowledgments}
\label{sec:ack}
The work of S. Park and J. Kang was supported by the National Research Foundation of Korea (NRF) grant funded by the Korea government (MSIT) (No. 2017R1A2B2012698). The work of O. Simeone was supported by the European Research Council (ERC) under the European Union's Horizon 2020 research and innovation programme (grant agreement No. 725731). 